\documentclass[journal]{IEEEtran}
\usepackage{amsmath,amsfonts}
\usepackage{algorithmic}
\usepackage{algorithm}
\usepackage{array}
\usepackage[caption=false,font=normalsize,labelfont=sf,textfont=sf]{subfig}
\usepackage{textcomp}
\usepackage{stfloats}
\usepackage{url}
\usepackage{verbatim}
\usepackage{graphicx}
\usepackage{cite}
\usepackage{xcolor}
\usepackage{ulem}
\usepackage{multirow}
\begin{document}

\title{Agile Orchestration at Will: An Entire Smart Service-Based Security Architecture Towards 6G}

\author{Zhuoran Duan,
        Guoshun Nan~\IEEEmembership{Member,~IEEE,}
        Rushan Li,
        Zijun Wang,
        Lihua Xiong, 
        Chaoying Yuan,
        Guorong Liu,
        Hui Xu,
        Qimei Cui~\IEEEmembership{Senior Member,~IEEE,}
        Xiaofeng Tao~\IEEEmembership{Senior Member,~IEEE,}
        Tony Q.S. Quek~\IEEEmembership{Fellow,~IEEE}

\thanks{Zhuoran Duan, Guoshun Nan, Rushan Li, Zijun Wang, Lihua Xiong, Qimei Cui and Xiaofeng Tao
are with the National Engineering Research Center for Mobile Network
Technologies, Beijing University of Posts and Telecommunications (BUPT), Beijing
100876, China and BUPT Shenzhen Institute, Shenzhen, China. (\textit{Corresponding author: Guoshun Nan}).
 
{Chaoying Yuan and Guorong Liu are with the China Telecom Corporation Limited Research Institute.}

{Hui Xu is at Datang Mobile Communication Equipment Co. LTD.}

Tony Q.S. Quek is the Cheng Tsang Man Chair Professor with the Singapore University of Technology and Design (SUTD).

}

}

\maketitle

\begin{abstract}
The upcoming 6G will fundamentally reshape mobile networks beyond communications, unlocking a multitude of applications that were once considered unimaginable. Meanwhile, security and resilience are especially highlighted in the 6G design principles. However, safeguarding 6G networks will be quite challenging due to various known and unknown threats from highly heterogeneous networks and diversified security requirements of distinct use cases, calling for a comprehensive re-design of security architecture. This motivates us to propose ES3A (Entire Smart Service-based Security Architecture), a novel security architecture for 6G networks. Specifically, we first discuss six high-level principles of our ES3A that include hierarchy, flexibility, scalability, resilience, endogeny, and trust and privacy. With these goals in mind, we then introduce three guidelines from a deployment perspective, envisioning our ES3A that offers service-based security, end-to-end protection, and smart security automation for 6G networks. Our architecture consists of three layers and three domains. It relies on a two-stage orchestration mechanism to tailor smart security strategies for customized protection in high-dynamic 6G networks, thereby addressing the aforementioned challenges. Finally, we prototype the proposed ES3A on a real-world radio system based on Software-Defined Radio (SDR). Experiments show the effectiveness of our ES3A. We also provide a case to show the superiority of our architecture.  
\end{abstract}


\section{Introduction}
\label{sectionIntro}
\begin{figure}[t!]
\centering
\includegraphics[width=0.5\textwidth,trim={10 530 80 0}]{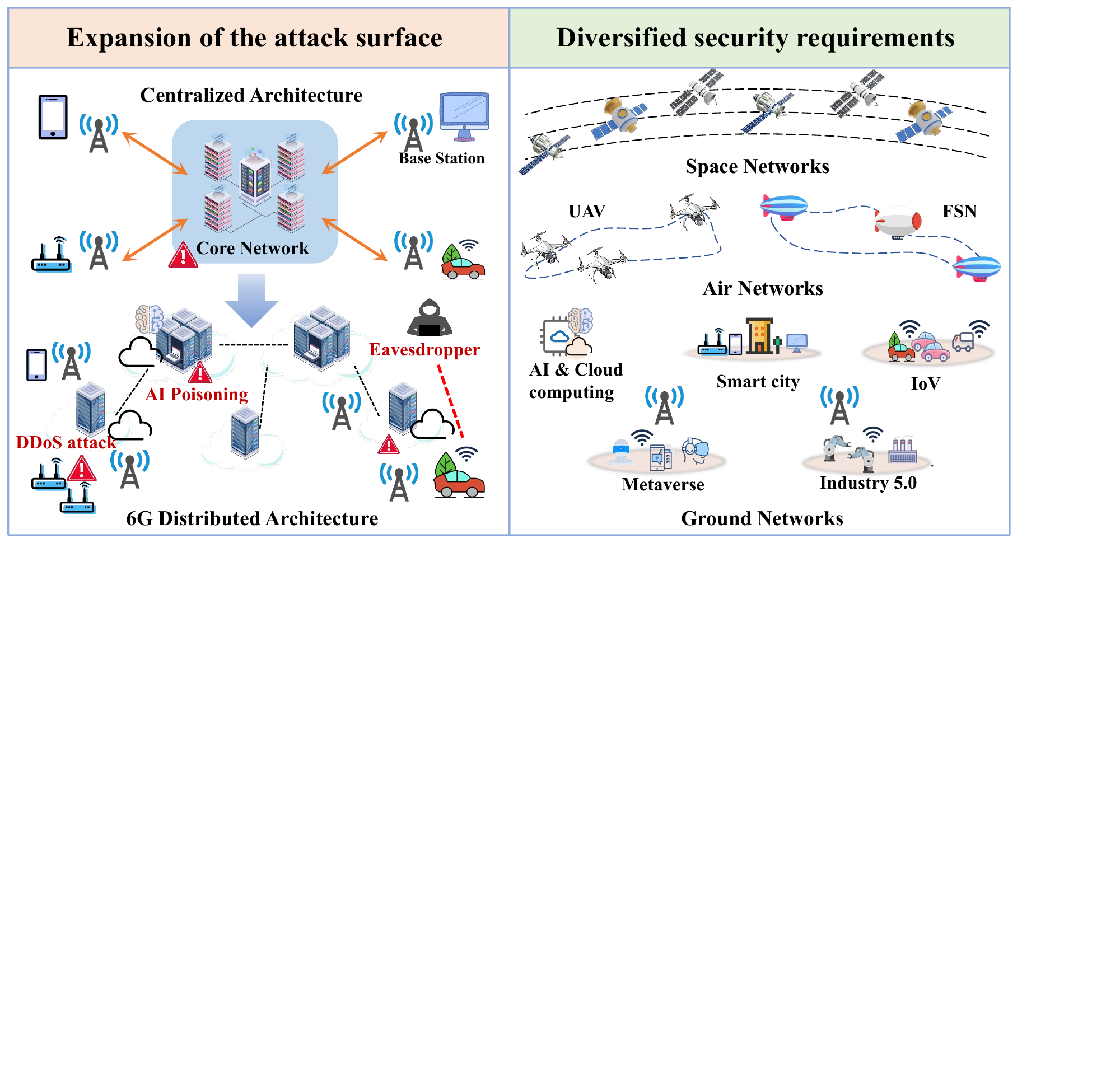}

\textcolor{black}{\caption{6G networks are evolving from centralized to distributed architectures, leading to an expansion of the attack surface. The diverse scenarios of 6G networks require different security needs. Addressing these issues necessitates a comprehensive redesign of the security architecture.}}
\label{fig-intro}
\vspace{-4mm}
\end{figure}

\IEEEPARstart{I}{nternational} Telecommunication Union (ITU) has officially marked the global vision of six-generation (6G) wireless networks, and the first commercial launch of 6G networks is expected around 2030. 6G will provide inherent intelligence, substantially higher capacity, broader coverage, and much lower latency, reshaping mobile networks beyond communications. The IMT-2023 framework~\cite{ref1} envisaged potential 6G use cases such as immersive communication, ubiquitous connectivity, AI and communication, facilitating various applications such as Industry 5.0, Internet of Vehicles (IoV), and digital twins. Figure \ref{fig-intro} shows space-air-ground integrated networks (SAGIN) in 6G, which integrates satellite systems, aerial networks, and terrestrial networks, offers ubiquitous coverage for numerous applications.

Security and resilience are especially highlighted in the design principles of the IMT-2030 6G vision, ensuring that networks are capable of surviving and quickly recovering from various threats or failures~\cite{ref1}. However, protecting 6G networks is quite challenging due to two preliminary reasons:
1) \textit{Increased Attack Surface:} 
Figure \ref{fig-intro} illustrates the transition of 6G networks from centralized to highly distributed architectures, dispersing core network functions across edge nodes, clouds, and virtualized infrastructures. This decentralization significantly expands the attack surface, exposing the network to threats such as unauthorized access and eavesdropping in wireless channels for the Internet of Vehicles (IoV), privacy attacks in Industry 5.0, and data poisoning in intelligent services. Additionally, unknown threats such as vulnerabilities in 6G protocols and quantum computing attacks pose potential risks. 2) \textit{Diverse Security Requirements:} Figure \ref{fig-intro} shows that 6G networks provide comprehensive coverage spanning space, air, and ground, serving a variety of application scenarios. These use cases have diverse security requirements. For example, the IoV and Industry 5.0 demand security services that meet low-latency requirements, while the vast number of sensors in smart cities require lightweight and highly efficient security solutions.

Existing efforts in 6G security architectures primarily consist of three categories: Zero-Trust Architectures~\cite{Zero-trust} (ZTA), Distributed Trust Management~\cite{DTM} (DTM), and Security Orchestration, Automation, and Response~\cite{soar} (SOAR), as summarized in Table~\ref{table}. ZTA addresses trust threats by implementing continuous identity verification and dynamic least-privilege access control, establishing a fine-grained trust model for 6G networks. DTM constructs a decentralized trust infrastructure for 6G, ensuring traceability and immutability of cross-domain entity behaviors through distributed ledger systems. While ZTA and DTM collectively provide 6G with adaptive and persistent trust frameworks. SOAR enhances real-time security control by creating closed-loop workflows that integrate threat detection, incident response, and remediation processes, thereby significantly accelerating threat response capabilities for 6G. Existing studies\cite{ref15,Huawei,DTM,Zero-trust,soar,HeXa-6G,NextG,JSAC,Mag} based on the three architectural paradigms outlined in recent white papers have significantly advanced both theoretical and practical aspects of 6G security.

\begin{table*}[t]
\label{table}
\centering
\caption{Comparison of Our Proposed ES3A and Existing Security Architectures}
\resizebox{1\textwidth}{!}{
\renewcommand{\arraystretch}{1.7}
\begin{tabular}{|>{\centering\arraybackslash}m{2.2cm}|>{\centering\arraybackslash}m{8cm}|*{5}{>{\centering\arraybackslash}m{2cm}|}}
\hline
\multirow{1}{*}{\large\raisebox{-4.8ex}{Architectures}} & 
\multirow{2}{*}{\large\raisebox{-4.8ex}{Key Features}} & 
\multicolumn{5}{c|}{\large Advantages} \\ \cline{3-7} 
 &  & \large Trust Management & \large Continuous Certification & \large Dynamic Adjustment & \large Closed-Loop Control & \large Customized Security \\ \hline
\large ZTA & \large Enforcing continuous identity verification and least-privilege access. & $\checkmark$ & $\checkmark$ & $\checkmark$ & $\times$ & $\times$ \\ \hline
\large DTM & \large Establishing tamper-proof credibility records via distributed ledger technology. & $\checkmark$ & $\checkmark$ & $\checkmark$ & $\times$ & $\times$ \\ \hline
\large SOAR & \large Automating workflows of orchestrating detection, investigation, and response. & $\times$ & $\times$ & $\checkmark$ & $\checkmark$ & $\times$ \\ \hline
\large ES3A & \large Achieving customized protection through distributed multi-domain collaboration. & $\checkmark$ & $\checkmark$ & $\checkmark$ & $\checkmark$ & $\checkmark$ \\ \hline
\end{tabular}
}
\end{table*}

However existing studies lack fine-grained discussion on security service customization. A critical challenge lies in providing customized security protection for 6G through security orchestration in distributed network environments. Enlightened by the above studies, we take a step further to propose ES3A, a novel security architecture designed for 6G networks, aiming to address the challenges of the highly distributed 6G networks and the customized security demands of various scenarios. ES3A establishes a distributed trusted environment through hierarchical design, while employing AI-driven orchestration to deliver customized security services. It prioritizes smart inter-domain collaboration to strengthen continuous certification and closed-loop control within distributed trust framework, highlighting customized security as shown in Table~\ref{table}.

In this paper, we first discuss the six high-level design principles of the proposed ES3A which include hierarchy, flexibility, scalability, resilience, endogeny, and trust and privacy. Keeping these principles in mind, we introduce three guidelines from a practical deployment perspective, including service-based security, end-to-end protection, and smart security automation. Equipped with these guidelines, we present our ES3A architecture and rely on a two-stage orchestration mechanism to yield smart protection policies. An \textit{AI-based policy agent} generates an orchestration policy in Stage 1, and a \textit{security automation manager} in Stage 2 forwards this policy for protection execution. We implement the proposed ES3A on a real-world radio system based on Software-Defined Radio (SDR). Experiments show the effectiveness of our  ES3A for agile security orchestration of 6G networks. The contributions of this paper are summarized as follows.

\begin{itemize}
    \item We present ES3A, an entire smart service-based security architecture for 6G networks. The proposed ES3A facilitates agile security orchestration tailored for 6G networks, addressing the challenges caused by network threats and diverse security requirements. 
    \item We introduce a two-stage service orchestration algorithm, where Stage 1 generates security policy and Stage 2 dynamically adjusts through inter-domain collaboration to optimize policy performance.
    \item We prototype the proposed ES3A on a real-world radio system and conduct extensive experiments. We provide a case study to show how our ES3A empowers the protection of IoT scenarios with agile security orchestration. 
\end{itemize}

\section{High-level Design Principles of our ES3A}
\label{section-principle}

\begin{figure*}[!t]
\centering
\includegraphics[width=1.0\textwidth,trim={15 280 15 250}]{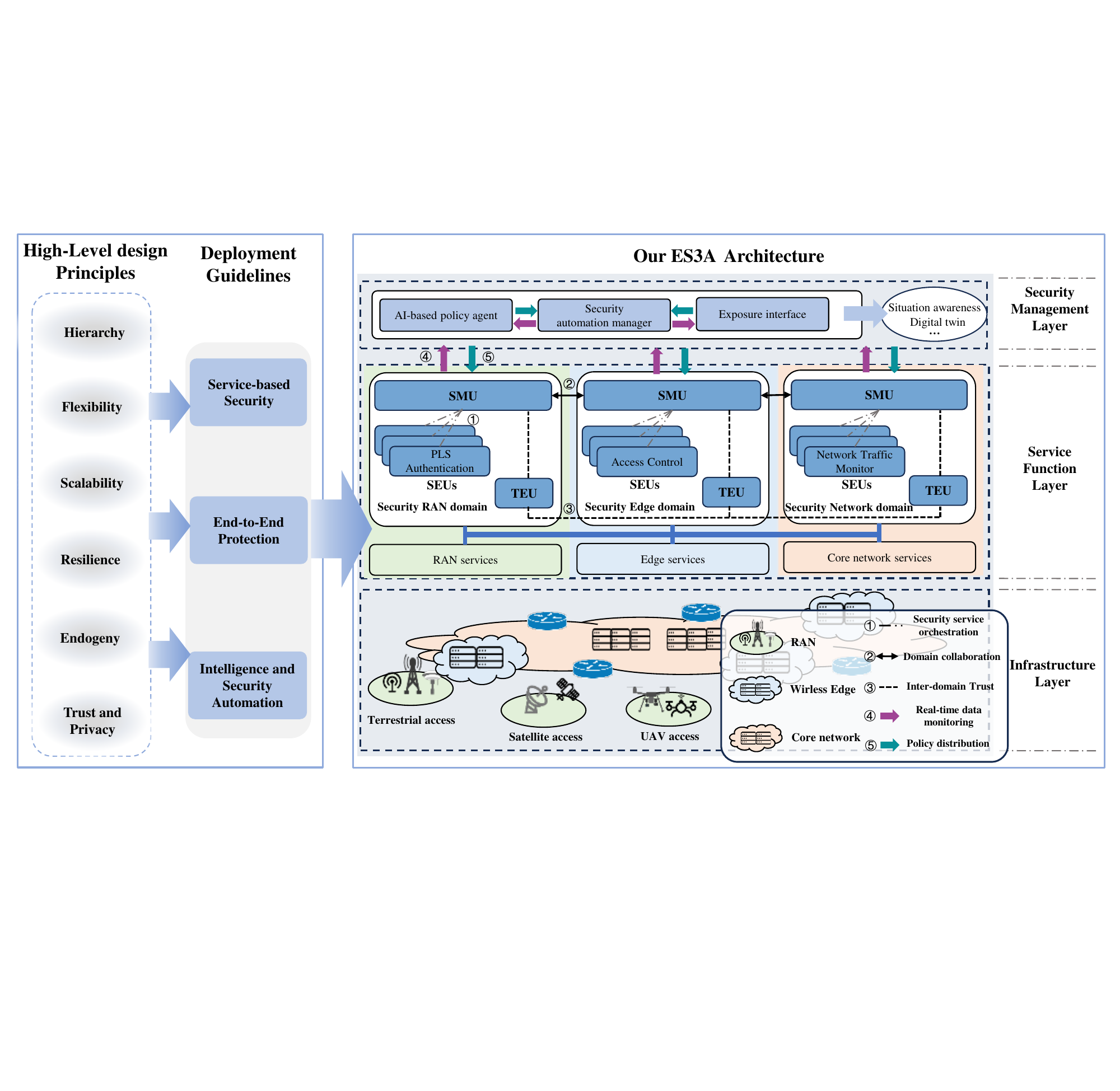}
\caption{Overview of our ES3A. The proposed ES3A is built on six high-level principles and three deployment guidelines. It consists of three layers, including the infrastructure layer, service function layer, and security management layer, and three domains, including RAN, edge, and core networks. 1) The management layer acts as a security engine to produce optimal security strategies using machine learning techniques. 2) The service function layer involves SMUs, SEUs, and TEUs to facilitate security deployment. 3) The infrastructure layer aims to virtualize the hardware and operation system resources for security services.}
\vspace{-3mm}
\label{fig:arc}
\end{figure*}

In this section, we summarize and elaborate on the rationale behind high-level design principles of ES3A and then show how we achieve these goals from a deployment perspective. 
The left side of Figure \ref{fig:arc} shows these principles and guidelines. 

\subsection{High-level Design Principles}
\noindent

\noindent
\textbf{Hierarchy:} 6G networks are highly decentralized, involving massive devices and subnetworks. For example, space-air-ground integrated networks in 6G include satellite systems, aerial networks, and terrestrial networks. A smart city deployment may involve millions of distributed sensors, each requiring secure communication and periodic updates. Thus, the security architecture should be hierarchical to align decentralized communication infrastructures of 6G networks.

\noindent
\textbf{Flexibility:} 6G networks will operate in a highly dynamic and heterogeneous environments, and may suffer from various known and unknown threats due to the expansion of attack surfaces. For example, satellite networks may be vulnerable to signal jamming or spoofing, while terrestrial networks could face DDoS attacks or other unknown ones. An attacker could manipulate training data or launch adversarial attacks to degrade the AI-based intrusion detection system. Therefore, the security architecture should adapt to these changes without requiring a complete overhaul.

\noindent
\textbf{Scalability:}
6G networks will connect billions of devices and are expected to adapt to the security demands of various emerging cases. For example, the number of cars in a vehicular network or sensors in IoT networks may rapidly increase, requiring scalable authentication and key management.
The security architecture should be able to scale to handle these dynamics while maintaining continuous protection.

\noindent
\textbf{Resilience:}
Resilience refers to the ability to continue to provide security services during and after disturbances~\cite{ref1}. It is a critical feature of the 6G security architecture due to the unprecedented complexity and dynamic nature of 6G networks. For example, if a base station fails to provide access service or faces attacks, the system can automatically reroute traffic and restore services using alternative nodes, ensuring uninterrupted services for users and applications. 

\noindent
\textbf{Endogeny:} Endogenous security indicates the concept of embedding security mechanisms directly into the design of 6G networks, rather than stacking them as external or afterthought measures. Thus, security will be inherent and proactive to various attacks and failures. For example, smart traffic systems will rely on various AI-driven services to secure real-time communications. Endogenous security embeds safeguards into the design of these AI processes, ensuring their robustness against adversarial attacks.

\noindent
\textbf{Trust and Privacy:}
6G will involve billions of IoT sensors, wearables, and autonomous systems, generating vast amounts of data. These high-sensitive data may face privacy attacks or be used for surveillance or malicious purposes. Meanwhile, safety-critical AI techniques are ubiquitous in 6G communications, requiring high-level data protection to ensure user privacy. Meanwhile, zero-trust protection can be activated based on service requirements. For example, self-driving cars rely on 6G for real-time communications, and threats to trust or privacy can lead to severe safety risks.

\subsection{Deployment Guidelines}
We follow the high-level design principles of 6G security to give three corresponding deployment guidelines as follows.

\noindent
\textbf{Service-based Security:}
6G networks are expected to employ service-based architecture (SBA) in RAN and core networks~\cite{ref4}. Compared with traditional monolithic architectures, SBA offers greater flexibility, scalability, and efficiency. Therefore, employing service-based security for 6G networks meets the high-level design principles. Equipped with network virtualization and cloud-based infrastructure, service-based 6G security can also benefit the system's resilience against various threats by dynamically scheduling various modules such as authentication, encryption, and intrusion detection. This deployment guideline can be seamlessly applied to the popular Open Radio Access Network (O-RAN)~\cite{O-RAN} to secure the system among different service providers.  

\noindent
\textbf{End-to-end Protection:}
End-to-end (E2E) security ensures that every component, communication link, and data flow within 6G networks can be well protected against various potential threats, vulnerabilities, and privacy attacks. Thus, it can mitigate the challenges of expansion of attack surface by managing protections across all layers, domains, and networks, thereby achieving the goals of hierarchy, flexibility, scalability, resilience, and trust and privacy. Such a design significantly differs from 5G E2E protection~\cite{ref2}, which emphasizes more on the security of communication links based on access authentication and data encryption. 

\noindent
\textbf{Intelligence and Security Automation:} Inherent intelligence aligns with the 6G use case integrated AI and communications, delivering learning-based security service to predict, detect, and mitigate risks. Security automation facilitates smart service orchestrations in both RAN and 6G core networks, automatically responding to incidents and adjusting security policies. Thus, it can greatly reduce the complexity of network security maintenance for 6G use cases such as ubiquitous connectivity and massive communication, providing flexible, scalable,  resilient, and trustworthy protections. Furthermore, service automation can also facilitate detecting threats and taking actions in real-time, thus benefiting 6G use cases such as hyper reliable and low-latency communication.

\begin{figure*}[!t]
\centering
\includegraphics[width=1.5\textwidth,trim={15 500 75 340}]{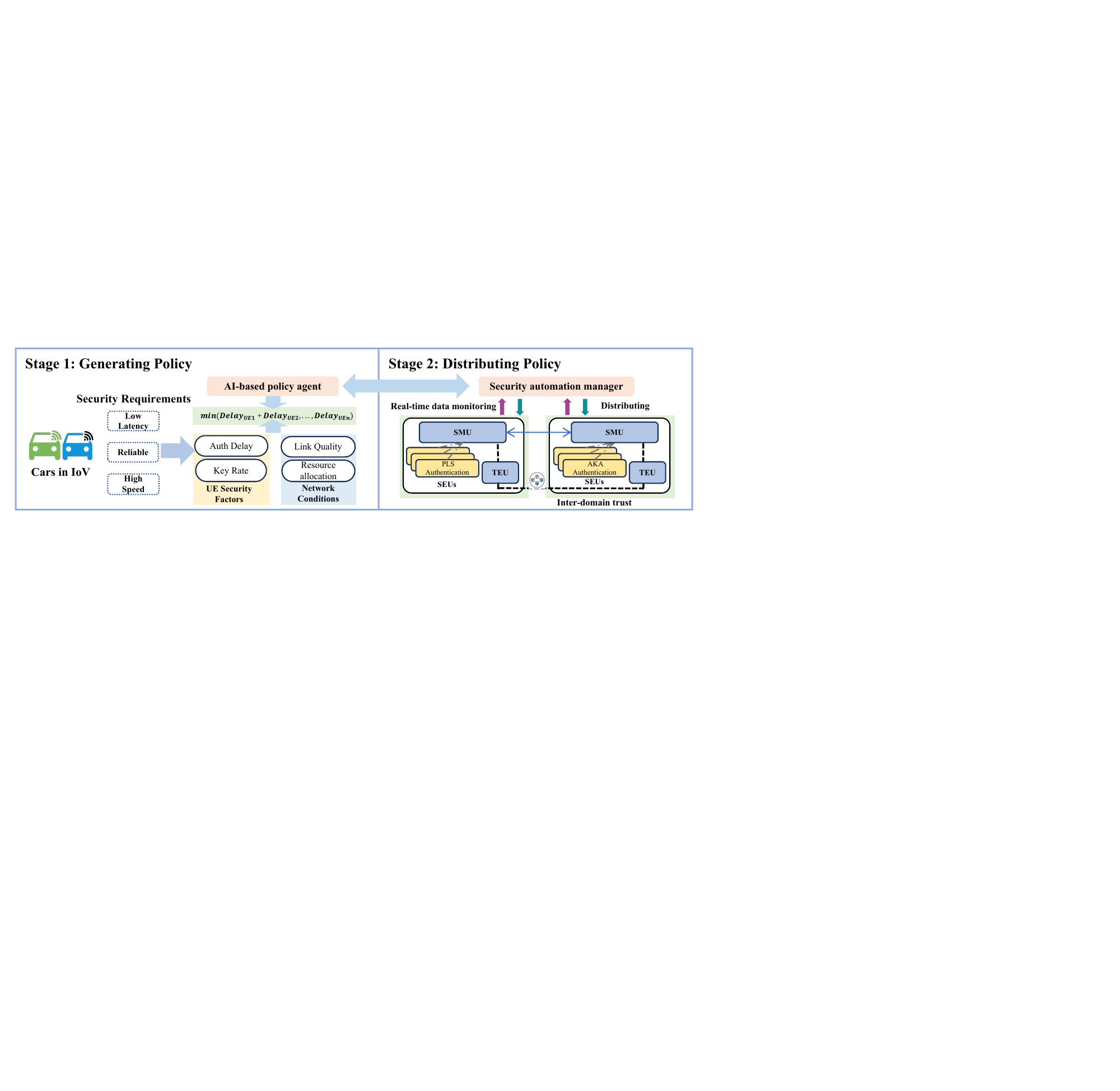}
\caption{Two-stage security orchestration of our ES3A of 6G networks. We rely on the AI-based policy agent to generate an orchestration policy in stage 1, where the agent takes security requirements, security factors, and network conditions as inputs for global optimizations. Then we feed the orchestration policy to the security automation manager in stage 2. The manager distributes the policy to SMUs for service execution to offer end-to-end protections. Meanwhile, the manager also responds to the policy agent in stage 1 for real-time interaction to obtain optimized orchestration.}
\label{fig:orchestration}
\vspace{-4mm}
\end{figure*}

\section{overall of the proposed ES3A architecture}
\label{sectinDesign}
Keeping the above deployment guidelines in mind, we propose an entire smart service-based security architecture (ES3A). The right side of Figure \ref{fig:arc} demonstrates our ES3A. It consists of three type of layers, including the infrastructure layer, service function layer, and security management layer. The service function layer of the proposed ES3A involves three different domains, including RAN domains, edge domains, and core network domains.  

\subsection{Security Layers: A Hierarchical Design for 6G Networks}

\noindent
\textbf{Infrastructure Layer:} It encompasses the physical equipment of the 6G networks. This layer relies on network virtualization to map physical devices in 6G networks such as base stations, routers, and servers, into logical resources. These resources, including storage, computation, and AI models, are managed via a unified resource management module to interact with upper-layer security services. 

\noindent
\textbf{Service Function Layer:} 
This layer encompasses various functions, services, and capabilities required to facilitate the deployment guidelines of service-based security and end-to-end protection. Service function layers play a pivotal role in ensuring that all network services and functions are secure at the design stage, thus allowing the \textit{endogenous safety} of 6G networks. We follow the deployment guidelines to build three components, including the security enable unit (SEU), which focuses on building atomic security capabilities for fine-grained service orchestration; the trust enable unit (TEU), which serves as the credit anchors to facilitate trustworthy collaborations within network and across service operators; the security management unit (SMU), which coordinates with each other and oversees SEUs and TEUs to deliver end-to-end protection. The service function layer also involves three types of security domains, including RAN, edge, and core networks, to offer customized security services for various 6G use cases. We will detail it in Section \ref{sec:arc-domain}.

\noindent
\textbf{Security Management Layer:} This layer acts as an engine of the security system of 6G networks. It globally manages and automates security services, allowing security to be dynamically adapted to changing network conditions and user requirements. It mainly consists of three modules, including AI-based policy agent, which aims to generate global security policies corresponding to network conditions and user requirements; Security automation manager, which can distribute the security policies to SMU and interacts with various security domains to offer real-time monitoring and adaptations; And exposure interface that is responsible for securing data-driven applications such as digital twin and situation awareness. 

\subsection{Security Domains: Orchestration and Trust Management}
\label{sec:arc-domain}
Security domains in the service function layer refer to distinct functional or operational areas within 6G networks, providing a structured approach to modular management and attack isolation in heterogeneous communications. We introduce three types of domains in our ES3A, including RAN, edge, and core networks. In each domain, SEUs, TEUs and SMUs cooperate to build effective defense lines or offer tailored services via agile orchestration. For intra-domain interaction, SMU is responsible for the exchange of security context between domains, and TEU serves as the credit anchor to guarantee inter-domain trust. We can employ blockchain or digital signatures to form the credit anchor. The collaborations between domains can also benefit the declaration of unknown attacks by exploring the various cues in each domain. Such a design of security domains in our architecture also facilitates the zero-trust paradigm via dynamic trust evaluation and domain collaboration, thereby aligning with the zero-trust architectures~\cite{Huawei,Zero-trust} that have been studied in previous works.

\section{Two-stage Agile Orchestration}
Figure \ref{fig:orchestration} illustrates the workflow of our two-stage orchestration including policy generation and policy execution in the security management layer and service function layer. This smart orchestration relies on security components such as policy agent, automation manager, and SMU, to deliver customized security services in an end-to-end manner. 

\subsection{Stage 1: Generating Policy}
The policy agent of the proposed ES3A produces orchestration strategies based on user requirements and network conditions. The left side of Figure \ref{fig:orchestration} demonstrates that high-speed cars in the Internet of Vehicles (IoV) need ultra-low-latency and reliable services. Our AI-based Policy Agent takes the application requirements such as latency, reliability, security factors such as authentication delay, secret key rate, and network conditions such as link qualities and resource allocations, as inputs. The agent then generates the optimal orchestration policy using a machine learning method. Here, we develop a reinforcement model to produce policy iteratively. The agent's output will be fed into the security automation manager for policy execution.  

\subsection{Stage 2: Distributing Policy}
The right side of Figure \ref{fig:orchestration} illustrate the workflow of the stage 1. The security automation manager takes the orchestration policy generated in the first step as inputs, and forwards these policies to SMUs that are located in different domains including RAN, edge, and core networks. Then, SMUs coordinate atomic service measures in SEUs for security orchestration, and TEUs provide trustworthy collaborations across domains. Meanwhile, the security automation manager also timely responds to the policy agent in the stage 1 to help the agent take further actions. Thus, we can build customized security services for 6G networks in an end-to-end manner.

\section{Experiments}

\begin{figure*}[t!]
\centering
\includegraphics[width=1\textwidth,trim={0 480 20 40}]{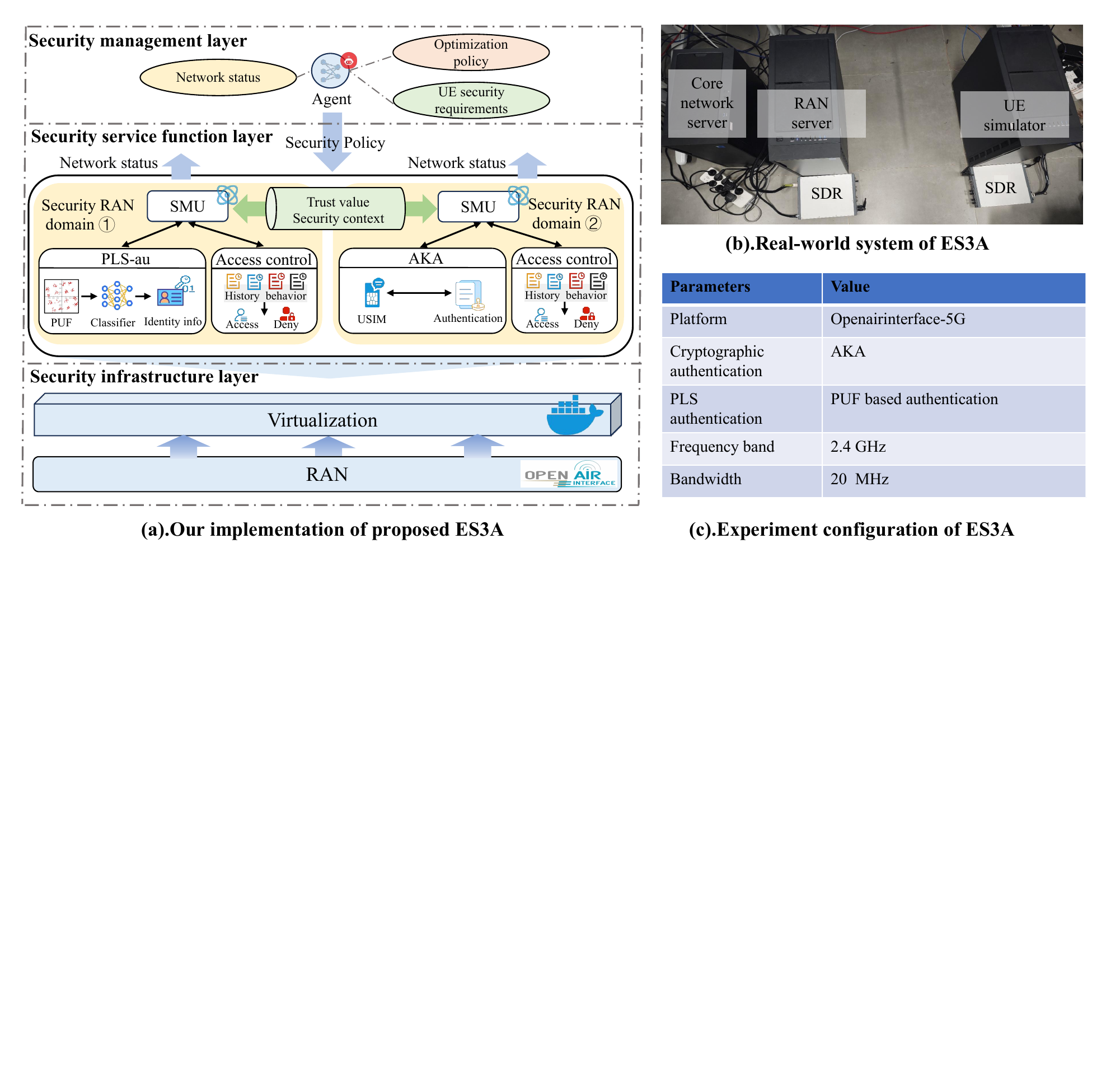}
\caption{Propotype and real-world system of our ES3A.}
\label{fig:real-system}
\vspace{-3mm}
\end{figure*}

\subsection{Implementation Details}

\noindent
\textbf{Implementation of the Infrastructure Layer}: 
We prototype the proposed ES3A in a real-world wireless system based on software-defined radio (SDR) and Openairinterface-5G (OAI)~\cite{OAI}. Figure \ref{fig:real-system} (a) demonstrates the detailed implementation of our prototype system in three layers and three domains. Figure \ref{fig:real-system} (b) illustrates the real-world system, which consists of three components, including a UE simulator with an SDR device, a base station (BS) with an SDR device, and core networks. The UE simulator relies on UERANSIM to configure mobile terminals~\cite{OAI}. Our system is deployed on three servers, each running Ubuntu 20.04 LTS and equipped with 64 GB RAM and an Intel Xeon Silver 4210 CPU (10-core, 2.2 GHz). The SDRs operate at a center frequency of 2.4 GHz with a fixed bandwidth of 20 MHz. The RAN and core network rely on the implementation of 3GPP 5G standard protocols for communications. Meanwhile, we run each security component based on containers. 

\noindent
\textbf{Implementation of the Security Function Layer:} We manage SEUs by virtualizing them as Network Functions (NFs), where each SEU can be regarded as an NF providing atomic capabilities. The SMU can create a set of SEU instances based on the security domain configuration. We configured two security domains (SecRAN 1 and SecRAN 2) to illustrate the workflow of our ES3A. The security domains are located on the NAS layer at the RAN side and include two SEUs for authentication and access control. The authentication services employ both cryptographic 5G-AKA and physical layer authentication based on radio fingerprinting~\cite{ref2} to deliver customized security for diverse scenarios. The physical layer authentication uses the design from~\cite{pls} for functionality implementation. For access control, we implemented a trust assessment and packet filtering mechanism based on Bayesian inference~\cite{coll-ids}. The UE's trust value \( T_D \) in each domain \( D \) is shared across security domains, used to compute and update the current UE trust value \( T_{UE} \), and compared against a threshold \( T_{th} \) for access control.

\noindent
\textbf{Implementation of the Security Management Layer:} We construct a RL-based agent at the management layer to provide two-stage security orchestration process. First, the user initiates an identity authentication request after establishing wireless access. The agent formulates the user's security policy \( P_{UE} \) based on their security requirements, network status, and optimization strategies, providing security services within the security domain. Once authentication is completed and key distribution is finalized, the SEU performs real-time trust assessment and access control by domains collaboration to achieve closed-loop control. Additionally, the transfer of trust value and security context across domains ensures that the UE's security services can switch between security domains on the RAN side. This switching can be adaptively adjusted by the agent or reselected based on user requirements and management layer policies.

\begin{figure*}[t!]
\centering
\includegraphics[width=1\linewidth,trim={120 220 123 530}]{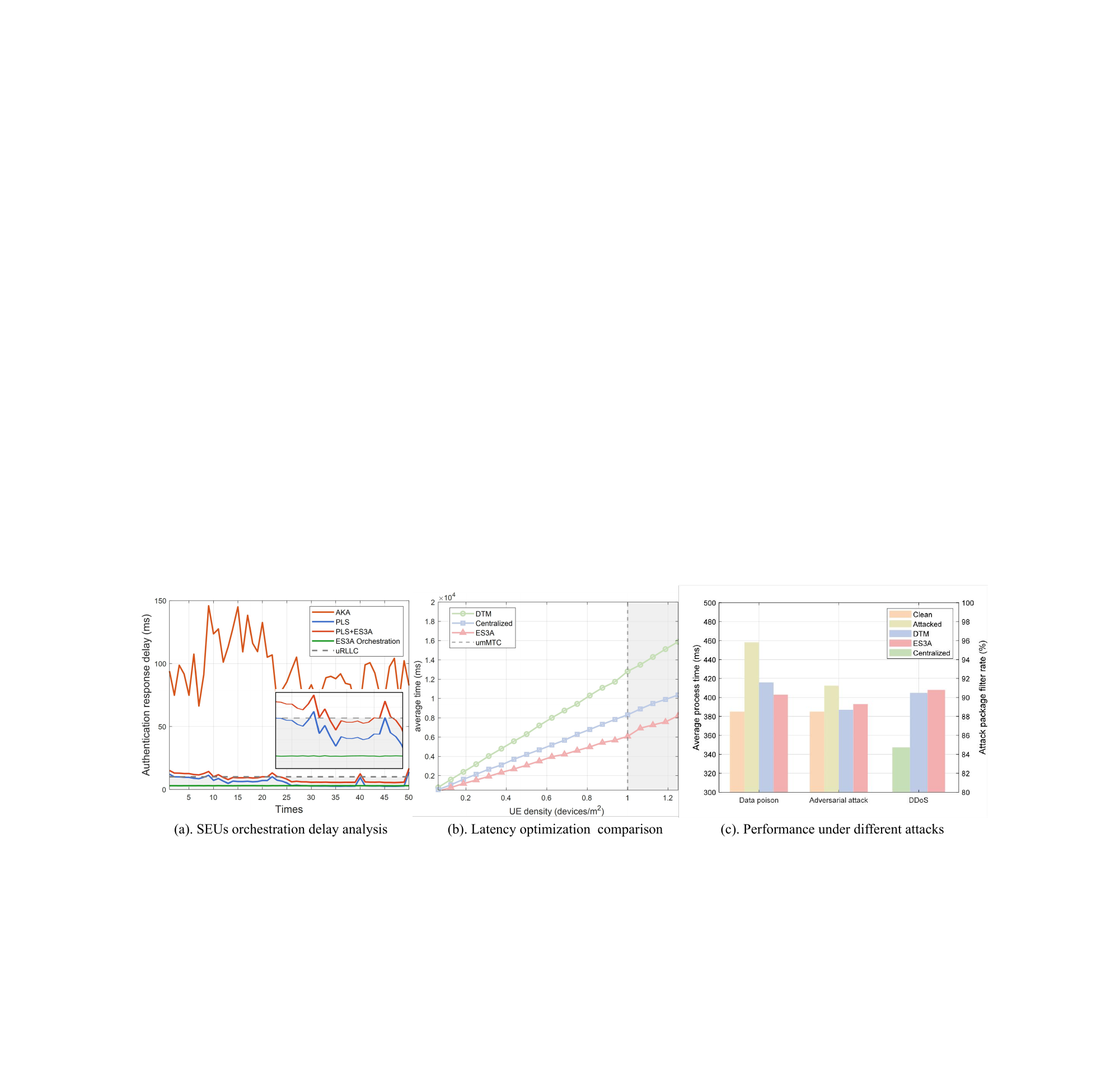}
\caption{Experimental Results in IoT Scenario.}
\label{fig:experiment}
\vspace{-3mm}
\end{figure*}

\subsection{Case study}

\textbf{Scenario Description:} 
We demonstrate the advantages of ES3A through an IoT case study. IoT is one of the key application scenarios for 6G, where a large number of sensors and mechanical devices require continuous network authentication. This necessitates the network to customize differentiated security solutions based on device types to balance security and low latency. We classify devices into two security levels: for large-scale IoT devices like sensors, lightweight physical layer authentication is used, while for UEs with high-security requirements, such as robots and factory machinery, robust cryptographic authentication is employed. Additionally, considering that physical layer authentication is constrained by channel conditions and device radio frequency fingerprinting, cryptographic authentication is adopted when physical layer authentication fails to meet the conditions. We demonstrate the performance advantages of the architecture through experiments on latency, network optimization and network attack.

We simulate 50 UEs on the system platform, with 40 nodes functioning as sensor devices using physical layer authentication and 10 nodes as industrial devices using cryptographic authentication. Meanwhile, we set up 3 BS nodes: BS1 and BS2 belong to SecRAN 1, while BS3 belongs to SecRAN 2. Both security domains adopt the same security service configuration as described in the previous chapter. These nodes randomly initiate network access requests to these BSs. For comparative evaluation, we established two baselines alongside ES3A: 1.Centralized architecture: UEs connect via nearest-neighbor access with rule-based security policies. 2.Distributed trust management (DTM): A distributed trust framework constructed using the method in~\cite{DTM}, employing agent-based security policy orchestration.

\textbf{Overhead Analysis:} The two SEUs provide differentiated security protections. The latency overhead of security service orchestration for a single UE under ES3A is shown in Figure~\ref{fig:experiment} (a). Compared to PLS, which is susceptible to channel conditions, AKA delivers robust authentication but introduces additional time overhead due to the need for core network identity verification. In contrast, PLS is more suitable for low-latency authentication. ES3A provides an agent-based orchestration, with an average process time overhead of 2.98 ms, meeting the uRLLC latency requirement of $\leq$ 10 ms. Experimental results demonstrate that ES3A can provide PLS services with an average latency of 8.6 ms for low-latency applications, along with near-real-time security orchestration.

\textbf{Experiments for Latency Optimization:} We validate the impact of management layer strategy optimization on network performance by analyzing the relationship between processing latency and UE scale across three SecRANs. The management layer agent selects the optimal authentication service and timing for different UEs to minimize the average authentication delay. In the simulation, we scaled the number of access requests per UE per iteration to emulate varying connection densities. As shown in Figure~\ref{fig:experiment} (b), both ES3A and DTM employ agent-based orchestration strategies to optimize the network's average latency overhead. However, ES3A demonstrates superior performance through inter-domain collaboration by offloading UE tasks across different security domains. Under umMTC connection density conditions, the experimental results show an average network authentication latency of 6 seconds. ES3A can further enhance network performance by scaling the number of security 
domains.

\textbf{Performance under Attacks:} We conduct experiments on three attack to evaluate the performance. We use the SIR model~\cite{SIR_model} to simulate device nodes being infected by worms. The infected UEs perform three types of threat: (i) DDoS on SecRAN to exhaust NFs resources, (ii) poisoning of training stage to degrade model accuracy, (iii) gradient estimation and adversarial attack FGSM perturbations during inference to cause mistake. Figure 5 (c) presents the performance under different attack. The centralized architecture demonstrates the worst performance against DDoS attacks, while both DTM and ES3A achieve higher malicious packet filtering rates through trust value-based access control. ES3A maintains robustness against AI-oriented attacks. Although compromised agents may lead to inaccurate orchestration strategies, the attack impact remains limited because agents comprehensively consider both network and UE states. Furthermore, the continuous access control combined with inter-domain collaboration effectively mitigates attacks, as malicious UEs must sustain continuous attack attempts to remain effective.

\section{Discussion and Future Directions}

\noindent
\textbf{Standardization Aligned Design}:
The ES3A design aligns with the ITU's vision toward 6G, while meeting differentiated security requirements. The ES3A prototype system implements critical modules such as the SMU in the form of Network Functions based on the SBA, maintaining compatibility with the existing 5G style. It enables smooth 6G evolution while avoiding additional protocol overhead. For the RAN side, ES3A requires orchestratable SEUs to support customizable security capabilities. Due to the limitations of the current 3GPP R15 DU-CU architecture, ES3A cannot directly comply with existing RAN protocols. Considering the service-oriented vision of 6G RAN, particularly with O-RAN paradigm~\cite{O-RAN} ES3A can deploy SMUs and SEUs within the Near-RT RIC to enable real-time security orchestration and AI-driven closed-loop control. As 6G standardization progresses, we will conduct further research to integrate ES3A with future network designs.

\noindent
\textbf{Data and AI Privacy}:
Data and AI privacy is one of the key challenges in 6G networks. It may trigger privacy leakage risks when ES3A suffers network intrusion. Security domains could become untrusted third parties due to attacks, increasing the risk of information exposure. TEU can serve as a trust anchor to provide inter-domain trust support. These risks can be mitigated through cross-domain challenge-response mechanisms and distributed trust management. Furthermore, for AI and other data-driven services, privacy protection can be achieved via homomorphic encryption and federated learning. These technologies can serve as SEUs to reinforce security.

\noindent
\textbf{Security Services Evaluation}: Designing customized security services for diverse 6G scenarios is one of the primary objectives of this architecture. Currently, there is a lack of discussion on the customization of security services, particularly in terms of quantifying the performance of different security services and developing strategies for their selection. How to granularly evaluate security services from various providers and orchestrate them according to specific scenarios will pose a significant challenge. In the future, there is an urgent need to design a comprehensive framework that characterizes the performance of multiple security services to address the diversified security challenges of 6G.

\noindent
\textbf{Capabilities Exposure}: The ES3A incorporates open interfaces in its design, enabling the architecture to transmit data to higher-level applications such as situational awareness and digital twins, thereby providing more intelligent security. Situational awareness in 6G security architectures includes threat detection, anomaly detection, visual representation, and real-time response functions. Digital twins create virtual models of 6G networks through mathematical modeling and real-time data, synchronizing them with their actual operational states. By integrating digital twins with situational awareness, security orchestration and management can be achieved with higher precision and in real time. However, building and maintaining these data-driven methods require substantial computational resources and storage, especially for complex 6G networks, making it crucial to reduce computational costs and complexity.

\section{Conclusion}

This article proposes ES3A, a novel security architecture for 6G networks, aiming to address security challenges under highly dynamic 6G networks and customized requirements. We discuss the six high-level design principles of 6G security architecture and then present three guidelines from a practical deployment perspective. Equipped with this guidance,  the proposed architecture involves three layers and three domains to achieve service-based security, end-to-end protection, and smart security automation. We rely on a two-stage orchestration mechanism to generate and execute agile security policies. Finally, we prototype the proposed ES3A on a real-world radio system and conduct experiments to show the effectiveness of our architecture. In the future, we plan to apply our ES3A to real-world large-scale networks.

\end{document}